\documentclass[sigconf]{acmart}

\usepackage{booktabs}     %

\usepackage{graphicx}
\usepackage{subcaption}
\usepackage{placeins} %
\usepackage{float}     %
\usepackage{newfloat}
\DeclareFloatingEnvironment[name=Listing, fileext=lol, placement=tbp, within=none]{listing}
\usepackage{tcolorbox}
\tcbuselibrary{skins,breakable,listings}
\usepackage{listings}
\usepackage{xcolor}  %
\usepackage{array}
\lstdefinestyle{code}{
  basicstyle=\ttfamily\small,
  numbers=left, numberstyle=\scriptsize, numbersep=6pt,
  breaklines=true, breakatwhitespace=true,
  frame=single, framerule=0.4pt,
  tabsize=2, showstringspaces=false
}
\lstdefinestyle{cpp}{style=code, language=C++}
\lstdefinestyle{json}{style=code, language=}
\lstdefinestyle{txt}{style=code, language=}

\newtcblisting{codecard}[2][]{%
  enhanced, breakable,
  colback=white, colframe=black!12,
  boxrule=0.6pt, arc=2mm, left=1mm, right=1mm, top=1mm, bottom=1mm,
  title={#2}, fonttitle=\bfseries\footnotesize,
  listing only, listing options={#1}
}

\usepackage{colortbl}
\definecolor{diffadd}{RGB}{225,251,231}   %
\definecolor{diffdel}{RGB}{255,233,232}   %
\lstdefinelanguage{Swift}{%
  morekeywords={func,let,var,while,for,in,if,else,guard,return,true,false,nil},
  sensitive=true, morestring=[b]", morecomment=[l]{//}}
\lstdefinestyle{swiftmini}{%
  basicstyle=\ttfamily\scriptsize, columns=fullflexible, keepspaces=true,
  language=Swift, keywordstyle=\bfseries\color{blue!60!black},
  literate={"}{{\textquotedbl}}1}
\newcommand{\gplus}{\textcolor{green!55!black}{\ttfamily\scriptsize +}}
\newcommand{\gminus}{\textcolor{red!70!black}{\ttfamily\scriptsize $-$}}
\newcommand{\dnum}[1]{\textcolor{black!40}{\ttfamily\scriptsize #1}}

\newcommand{\cd}[1]{{\ttfamily\scriptsize\strut #1}}
\newcommand{\kw}[1]{\textcolor{blue!60!black}{\textbf{#1}}}
\newcommand{\diffline}[4]{%
  \cellcolor{#1}\makebox[1.3em][r]{\dnum{#2}}\hspace{8pt}%
  \makebox[0.6em][c]{#3}\hspace{6pt}\cd{#4}}

\usepackage{tabularx}

\usepackage{wrapfig}
\usepackage{multirow}
\usepackage{multicol}
\usepackage{enumitem}
\usepackage{listings}
\usepackage{xspace}
\usepackage{subcaption}
\usepackage{pifont}

\usepackage{tikz}
\tikzset{>=latex}
\usetikzlibrary{shapes.geometric,arrows.meta,positioning,fit,backgrounds,calc} %

\tikzset{
  ic/.style={line width=0.6pt, line cap=round, line join=round},
  pics/play/.style={code={
    \draw[ic] (0,0) circle (0.26);
    \fill (-0.08,-0.13) -- (-0.08,0.13) -- (0.16,0) -- cycle;
  }},
  pics/search/.style={code={
    \draw[ic] (0.03,0.06) circle (0.16);
    \draw[ic] (-0.09,-0.06) -- (-0.22,-0.19);
  }},
  pics/bug/.style={code={
    \draw[ic] (0,0) ellipse (0.15 and 0.21);
    \draw[ic] (0,0.21) -- (0,0.30);
    \draw[ic] (0,0.30) circle (0.07);
    \draw[ic] (-0.04,0.36) -- (-0.09,0.43);
    \draw[ic] (0.04,0.36) -- (0.09,0.43);
    \foreach \y in {0.10,0,-0.10}{
      \draw[ic] (-0.13,\y) -- (-0.26,\y+0.05);
      \draw[ic] (0.13,\y) -- (0.26,\y+0.05);
    }
  }},
  pics/shield/.style={code={
    \draw[ic] (0,0.24) -- (-0.17,0.17) -- (-0.17,-0.07)
      .. controls (-0.17,-0.20) and (-0.07,-0.28) .. (0,-0.30)
      .. controls (0.07,-0.28) and (0.17,-0.20) .. (0.17,-0.07)
      -- (0.17,0.17) -- cycle;
    \draw[ic] (-0.08,0.0) -- (-0.02,-0.08) -- (0.10,0.10);
  }},
  pics/funnel/.style={code={
    \draw[ic] (-0.22,0.20) -- (0.22,0.20) -- (0.05,-0.05) -- (0.05,-0.22)
      -- (-0.05,-0.22) -- (-0.05,-0.05) -- cycle;
  }},
  pics/db/.style={code={
    \draw[ic] (-0.16,0.18) arc (180:360:0.16 and 0.07);
    \draw[ic] (-0.16,0.18) arc (180:0:0.16 and 0.07);
    \draw[ic] (-0.16,0.18) -- (-0.16,-0.18);
    \draw[ic] (0.16,0.18) -- (0.16,-0.18);
    \draw[ic] (-0.16,0.0) arc (180:360:0.16 and 0.07);
    \draw[ic] (-0.16,-0.18) arc (180:360:0.16 and 0.07);
  }},
  pics/brain/.style={code={
    \draw[ic] (0,0.26)
      .. controls (-0.22,0.26) and (-0.28,0.08) .. (-0.20,-0.02)
      .. controls (-0.30,-0.10) and (-0.20,-0.28) .. (-0.05,-0.22)
      .. controls (0.0,-0.30) and (0.0,-0.30) .. (0.05,-0.22)
      .. controls (0.20,-0.28) and (0.30,-0.10) .. (0.20,-0.02)
      .. controls (0.28,0.08) and (0.22,0.26) .. (0,0.26) -- cycle;
    \draw[ic] (0,0.26) -- (0,-0.24);
    \draw[ic] (-0.10,0.12) .. controls (-0.02,0.10) and (-0.02,0.0) .. (-0.10,-0.04);
    \draw[ic] (0.10,0.12) .. controls (0.02,0.10) and (0.02,0.0) .. (0.10,-0.04);
  }},
  pics/book/.style={code={
    \draw[ic] (0,-0.16) .. controls (-0.10,-0.24) and (-0.24,-0.20) .. (-0.26,-0.16)
      -- (-0.26,0.16) .. controls (-0.20,0.12) and (-0.06,0.10) .. (0,0.18);
    \draw[ic] (0,-0.16) .. controls (0.10,-0.24) and (0.24,-0.20) .. (0.26,-0.16)
      -- (0.26,0.16) .. controls (0.20,0.12) and (0.06,0.10) .. (0,0.18);
    \draw[ic] (0,0.18) -- (0,-0.16);
  }},
  pics/graph/.style={code={
    \draw[ic] (-0.30,0) -- (0.30,0);
    \draw[ic] (0,0) -- (0,0.20);
    \foreach \x in {-0.30,-0.10,0.10,0.30}{ \fill (\x,0) circle (0.035); \draw[ic] (\x,0) circle (0.035);}
    \fill (0,0.20) circle (0.035); \draw[ic] (0,0.20) circle (0.035);
  }},
  pics/repo/.style={code={
    \draw[ic] (-0.10,0.26) -- (-0.10,-0.26);
    \foreach \y in {0.20,-0.04,-0.20}{ \fill (-0.10,\y) circle (0.05); \draw[ic](-0.10,\y) circle (0.05);}
    \draw[ic] (-0.10,-0.04) .. controls (0.06,-0.04) and (0.16,0.04) .. (0.16,0.16);
    \fill (0.16,0.20) circle (0.05); \draw[ic] (0.16,0.20) circle (0.05);
  }},
}

\newcommand*\circled[1]{\tikz[baseline=(char.base)]{
            \node[shape=circle,draw,fill=black,text=white,inner
            sep=0.4pt] (char) {\small #1};}}

\usepackage[framemethod=tikz]{mdframed}
\mdfdefinestyle{mpdframe}{
   frametitlebackgroundcolor   =black!15,
    frametitlerule              =true,
    roundcorner                 =3pt,
    middlelinewidth             =1pt,
    skipabove                   =\topskip,
    skipbelow                   =\topskip,
    innermargin                 =0.1cm, %
    outermargin                 =0.1cm,
    innerleftmargin             =0.1cm,
    innerrightmargin            =0cm,
    innertopmargin              =0.1cm,
    innerbottommargin           =0.1cm,
    align=center   
}

\newcommand{\DefMacro}[2]{%
   \expandafter\newcommand\csname rmk-#1\endcsname{#2}%
}
\newcommand{\UseMacro}[1]{\csname rmk-#1\endcsname}

\newcommand{\Comment}[1]{}

\newcommand{\Space}[1]{}

\newcommand{\Code}[1]{{\small\ifmmode{\texttt{#1}}\else$\texttt{#1}$\fi}}
\newcommand{\CodeIn}[1]{{\small\ifmmode{\mathtt{#1}}\else$\mathtt{#1}$\fi}}
\newcommand{\ColorBack}[1]{%
  \begingroup \setlength{\fboxsep}{0pt}%
}

\definecolor{gray}{RGB}{211,211,211}

\newcommand{\PaperTitle}{The Vocabulary of Flaky Tests in Swift}

\newcommand{\RQOnea}{With what accuracy can flaky test cases be predicted from the vocabulary (identifiers and keywords) present in the source code of Swift test cases?}
\newcommand{\RQOneb}{Does the predictive performance achieved by vocabulary-based classifiers remain stable when flaky tests are labeled exclusively through local re-execution, without the commit-history mining component of the hybrid strategy?}
\newcommand{\RQTwo}{Does vocabulary-based prediction substantially outperform trivial baseline classifiers that disregard the test vocabulary?}
\newcommand{\RQThree}{Which identifiers present in Swift test code are most strongly associated with test instability?}
\newcommand{\RQFour}{When and why does the vocabulary-based classifier mispredict, and what do these errors reveal about the limits of lexical prediction?}

\newcommand{\x}{x\xspace}

\newcommand{\EQ}[1]{RQ#1\xspace}

\newcommand{\ourrepo}{\url{https://doi.org/10.5281/zenodo.22118906}}

\newcommand{\todo}[2][normal]{%
  \ifdefined\showtodos
    \ifstrequal{#1}{high}{\sethlcolor{red}}{}%
    \ifstrequal{#1}{normal}{\sethlcolor{yellow}}{}%
    \ifstrequal{#1}{low}{\sethlcolor{green}}{}%
    {\hl{\textbf{TODO:} #2}}%
    \marginpar{\textcolor{red}{\textbf{TODO}}}%
  \fi
}

\newif\ifshowtodos
\showtodostrue  %

\DefMacro{eq-single-performance}{\EQ{1}}
\DefMacro{eq-single-safety}{\EQ{2}}
\DefMacro{eq-evolution-stability}{\EQ{3}}
\DefMacro{eq-evolution-performance}{\EQ{4}}
\DefMacro{eq-evolution-safety}{\EQ{5}}
\DefMacro{Category}{Category}
\DefMacro{PolicyType}{Policy Type}
\DefMacro{Description}{Description}

\definecolor{SubtleColor}{rgb}{0,0,.50}

\definecolor{DefensiveColor}{RGB}{65,105,225}      %
\definecolor{PortableColor}{RGB}{34,139,34}        %
\definecolor{NormColor}{RGB}{205,92,92}            %
\definecolor{EnvColor}{RGB}{138,43,226}            %

\definecolor{UniqueSymptomColor}{RGB}{200,230,255}     %
\definecolor{ContextSymptomColor}{RGB}{255,250,205}    %

\AtBeginDocument{%
  }

\setcopyright{acmlicensed}
\copyrightyear{2026}
\acmYear{2026}
\acmDOI{XXXXXXX.XXXXXXX}
\acmConference[SAST 2026]{11th Brazilian Symposium on Systematic and Automated Software Testing}{September 8--11, 2026}{São Paulo, SP, Brazil}
\acmISBN{978-1-4503-XXXX-X/2018/06}

\setcopyright{none}
\renewcommand\footnotetextcopyrightpermission[1]{}

\usepackage{graphicx}
\usepackage{balance}
\usepackage{listings}

\usepackage{tcolorbox}

\tcbset{answerbox/.style={%
  fontupper=\normalsize,
  colback=black!6,
  boxrule=0pt, sharp corners, boxsep=2pt,
  left=2pt, right=2pt, top=2.5pt, bottom=2pt,
  before skip=2pt, after skip=2pt,}}

\usepackage{mdframed}
\usepackage{multirow}
\usepackage{fontawesome}
\usepackage{placeins}
\usepackage{balance}
\usepackage{hyperref}

\usepackage{pifont}%

\definecolor{codebg}{rgb}{0.95, 0.95, 0.92} %
\definecolor{keywordcolor}{rgb}{0.0, 0.0, 0.6} %
\definecolor{stringcolor}{rgb}{0.58, 0.0, 0.82} %
\definecolor{commentcolor}{rgb}{0.0, 0.5, 0.0} %

\newcommand{\myparagraph}[1]{\vspace{0.5ex}\noindent \textbf{#1. }}

\begin{document}

\title{\PaperTitle}

\author{João Medeiros}
\authornote{These authors contributed equally to this work.}
\affiliation{%
  \institution{Centro de Informática\\Universidade Federal de Pernambuco}
 \city{Recife}
 \state{Pernambuco}
 \country{Brazil}
}
\email{jpcm2@cin.ufpe.br}

\author{Denini Silva}
\authornotemark[1]
\affiliation{%
  \institution{Universidade Federal Rural de Pernambuco}
  \city{Belo Jardim}
  \state{Pernambuco}
  \country{Brazil}
}
\email{denini.gabriel@ufrpe.br}

\author{Breno Miranda}
\affiliation{%
    \institution{Centro de Informática\\Universidade Federal de Pernambuco}
  \city{Recife}
  \state{Pernambuco}
  \country{Brazil}
}
\email{bafm@cin.ufpe.br}
\renewcommand{\shortauthors}{Medeiros et al.}

\begin{abstract}
Flaky tests produce non-deterministic outcomes without code change, eroding CI confidence and delaying deliveries. While vocabulary-based machine learning prediction has proven effective for Java and JavaScript, no study has evaluated it for Swift, a language whose testing style is dominated by UI and asynchronous code.
We collect 91 flaky and 22,349 stable tests from 15 open-source Swift projects via re-execution and commit-history mining, then train five classifiers (\textit{Random Forest}, \textit{Decision Tree}, \textit{Naive Bayes}, \textit{SVM}, \textit{KNN}) on \textit{TF-IDF} unigram+bigram features under stratified 5-fold cross-validation.
\textit{Random Forest} achieves the best performance (Precision\,=\,0.92, F1\,=\,0.86, AUC\,=\,0.95) and substantially outperforms trivial baselines, among them a vocabulary-threshold rule applied to the most informative tokens, confirming a genuine discriminative signal (MCC\,=\,0.75 vs.\ 0.08 for the best baseline). Information-gain analysis reveals two complementary signal types. \emph{Flakiness markers} appear predominantly in unstable tests and comprise concurrency primitives (\textit{async}, \textit{await}), expectation-based synchronisation (\textit{expectation}, \textit{fulfill}), error propagation (\textit{throws}), and explicit timing dependence (\textit{timeout}, \textit{wait}, \textit{now}). \emph{Stability markers}, chiefly the assertion vocabulary of plainly synchronous tests (\textit{xctassertequal}), count as evidence \emph{against} flakiness. Error analysis shows that the model fails when flakiness is hidden in shared infrastructure outside the test body or when async constructs are used in a deterministic context, exposing the intrinsic limit of lexical prediction.
These results extend vocabulary-based flakiness detection to the Swift ecosystem and characterise both its effectiveness and its boundaries.
\end{abstract}


\maketitle

\section{Introduction}
\label{sec:introduction}
In software development, tests are fundamental for identifying defects and verifying that fixes work correctly \cite{Kapfhammer2004}. Ideally, test outcomes remain stable across executions of an unchanged codebase. However, \textit{flaky tests}, that is, tests that non-deterministically pass or fail without any change to the code or the execution environment, are a frequent reality \cite{miranda2020}. Their causes are diverse, including concurrency issues, asynchronous operations, test-order dependencies, mismanagement of external resources, dependence on system time, and the use of random values \cite{luo2014}, as well as restricted ranges of expected values and \textit{timeout}-related issues \cite{eck2019}.

Although flaky tests do not directly affect software functionality, they impose a substantial cost on deployment pipelines \cite{luo2014, herzig2015, micco2017}. \citet{micco2017} reported that 16\% of the 4.2 million tests evaluated in 2017 at Google exhibited unstable behavior, causing recurring delays in \textit{releases}, and that between 2\% and 16\% of computational capacity was consumed exclusively on re-executing these flaky tests.

Detecting flaky tests is itself challenging. The most common approach, repeated execution of the test suite \cite{bell2018, winter2020}, is costly and time-consuming. This has motivated alternative techniques based on coverage differences across versions, randomization of implementations, injection of stress on computational resources, and static analysis of code features associated with known root causes, including concurrency, asynchrony, and randomness \cite{parry2021}.

Among static approaches, machine learning models trained on test-code features have shown promise as predictors of instability \cite{miranda2021, alshammari2021}. \citet{miranda2020} introduced the notion that flaky tests exhibit a characteristic \textit{vocabulary}, that is, identifiers and keywords that occur disproportionately in them, and showed that classifiers such as \textit{Random Forest} and \textit{SVM} trained on this vocabulary reach an \textit{F-score} of 0.95 for Java. \citet{soratto2023} replicated this methodology for JavaScript, and \citet{ahmad2025} compared the linguistic diversity of flaky-test vocabularies across Java, Python, C++, Go, and JavaScript using a similar set of classifiers.

Despite this growing body of evidence, the vocabulary-based approach has, to the best of our knowledge, never been evaluated for Swift, the primary language for iOS, macOS, and other Apple-platform development. Swift differs from the ecosystems studied so far in ways that are directly relevant to flakiness: its concurrency model is centered on \textit{async}/\textit{await}, \textit{Grand Central Dispatch}, and actors, and its test suites rely heavily on UI tests written with \textit{XCTest}/\textit{XCUITest}, which interact with rendered interface elements rather than purely programmatic state. Empirical studies of flaky tests in mobile and UI-driven suites report root causes, such as UI-element timing and platform-specific concurrency, that are largely absent from the Java and JavaScript projects studied previously \cite{thorve2018, romano2021}. If the lexical patterns that signal instability in those ecosystems do not transfer to Swift, vocabulary-based tools built around them would offer limited help to iOS developers. This is the gap this work addresses: whether the unstable vocabulary identified for Java and JavaScript generalizes to a language and ecosystem with a markedly different concurrency model and a far stronger reliance on UI testing.

To address this gap, this work investigates the static prediction of \textit{flaky tests} in Swift using machine learning applied to the vocabulary extracted from the source code of the tests. Our approach is organized as a six-stage workflow (Section~\ref{sec:approach}) that (i) collects test cases from open source Swift projects through two complementary sources, namely systematic re-execution of test \textit{suites} and the mining of commit and pull-request histories where developers document and fix instability; (ii) labels each test and reduces it to its discriminative lexical units; and (iii) trains classification models on the resulting vocabulary to predict a test's propensity to be \textit{flaky} without executing it, while extracting an interpretable ranking of the tokens most strongly associated with instability.

We apply this methodology to a dataset of 91 flaky and 22,349 non-flaky tests collected from 15 open source Swift projects. Among the five classifiers evaluated (\textit{Random Forest}, \textit{Decision Tree}, \textit{Naive Bayes}, \textit{SVM}, and \textit{KNN}), \textit{Random Forest} achieved the best performance, with a Precision of 0.92, \textit{F1-Score} of 0.86, and \textit{AUC} of 0.95, with \textit{SVM} statistically indistinguishable from it. Information-gain analysis further shows that the most discriminative tokens divide into two complementary groups. \emph{Flakiness markers} appear predominantly in unstable tests and comprise concurrency primitives (\textit{async}, \textit{await}), expectation-based synchronisation (\textit{expectation}, \textit{fulfill}), error-propagation constructs (\textit{throws}), and explicit timing dependence (\textit{timeout}, \textit{wait}). \emph{Stability markers}, in turn, suppress the flaky prediction; the most robust of them is \textit{xctassertequal}, the assertion vocabulary of plainly synchronous tests.

The main contributions of this work are:
\begin{enumerate}
    \item We construct the first dataset of flaky and non-flaky Swift tests, combining systematic re-execution with the mining of commits and pull requests across 15 open source projects.
    \item We evaluate five machine learning classifiers for static, vocabulary-based flaky-test prediction in Swift, showing that an approach validated for Java and JavaScript transfers to a language with a distinct concurrency model and a strong reliance on UI testing.
    \item We identify a \textit{Swift vocabulary of flakiness}, showing that it comprises both \emph{flakiness markers} and previously unreported \emph{stability markers}, and relate both groups to root causes of instability reported for other languages.
    \item We structure the evaluation around four research questions that, beyond predictive accuracy, contextualize the classifiers against trivial baselines and qualitatively analyze representative misclassifications to delineate the limits of vocabulary-based prediction.
    \item We assess the robustness of the hybrid labeling strategy through a validation analysis restricted to tests confirmed as flaky via local re-execution.
\end{enumerate}

\section{Motivating Example}
\label{sec:motivating-example}

To ground the discussion that follows, consider \textit{testShareSheetSendToDevice}, a \textit{flaky} test from the \textit{firefox-ios} project~\cite{firefoxProj} whose repair we recovered from a \textit{pull request}~\cite{firefoxios_pr27270}. Listing~\ref{fig:motivating-diff} shows the test before and after that fix. In its flaky form, the test opens the browser's share sheet, waits for the ``Send to Device'' button to appear, taps it once (line~3, marked $-$), and then asserts that the expected follow-up screen is reached (lines~10--11). This test is representative of the UI-driven, asynchronous code that dominates Swift/iOS suites, and of the async-assertion patterns our study identifies as the dominant flakiness signal in Section~\ref{sec:RQ3}.

\begin{listing}[h]
\centering
\renewcommand{\arraystretch}{1.0}\setlength{\tabcolsep}{3pt}
\setlength{\arrayrulewidth}{0.4pt}\setlength{\fboxsep}{6pt}%
\fbox{%
\begin{tabular}[t]{@{}l@{}}
\multicolumn{1}{@{}c@{}}{\footnotesize\textbf{Flaky version and fix (PR \#27270)}}\\[2pt]\hline\noalign{\vskip2pt}
\diffline{white}{1}{}{\kw{func} testShareSheetSendToDevice() \{}\\
\diffline{white}{2}{}{~~openNewShareSheet()}\\
\diffline{diffdel}{3}{\gminus}{~~app.staticTexts["Send to Device"].waitAndTap()}\\
\diffline{diffadd}{3}{\gplus}{~~\kw{let} btn = app.staticTexts["Send to Device"]}\\
\diffline{diffadd}{4}{\gplus}{~~\kw{var} attempts = 2}\\
\diffline{diffadd}{5}{\gplus}{~~\kw{while} btn.isVisible() \&\& attempts > 0 \{}\\
\diffline{diffadd}{6}{\gplus}{~~~~btn.waitAndTap()}\\
\diffline{diffadd}{7}{\gplus}{~~~~waitForNoExistence(btn)}\\
\diffline{diffadd}{8}{\gplus}{~~~~attempts -= 1}\\
\diffline{diffadd}{9}{\gplus}{~~\}}\\
\diffline{white}{10}{}{~~waitForElementsToExist([ ... ])}\\
\diffline{white}{11}{}{~~...doneButton.waitAndTap()}\\
\diffline{white}{12}{}{\}}\\
\end{tabular}}
\caption{Flaky UI test and its fix~\cite{firefoxios_pr27270}: the single tap ($-$) becomes a bounded retry ($+$).}
\label{fig:motivating-diff}
\end{listing}

The instability arises from a subtle gap between an element \emph{existing} and being ready to receive input. The call to \texttt{.waitAndTap()} blocks only until the button is present in the view hierarchy; it has no way of knowing whether the share sheet has finished its entry animation. Because that animation runs asynchronously, the single tap on line~3 sometimes arrives while the view is still mid-animation, before it has settled into its final, stable position, and the UI layer simply drops the tap, with no error or visible effect. When this happens the screen never advances and the assertions on lines~10--11 fail; on a marginally slower run the tap lands and the very same test passes. Two outcomes for one unchanged revision is the defining symptom of \textit{flakiness}~\cite{luo2014}, and races of exactly this kind, between a test's actions and the timing of UI rendering, are among the most frequently reported causes of flakiness in mobile and UI-driven suites~\cite{romano2021, thorve2018, eck2019}.

The developer's fix (Listing~\ref{fig:motivating-diff}, lines~3--9) neither inserts a fixed sleep nor speeds the test up; instead, it makes the test \emph{observe} the effect of its own action. The tap is wrapped in a loop that re-taps while the button is still visible and calls \texttt{waitForNoExistence} (line~7) to confirm that the button has actually disappeared and the tap therefore took effect, capping the attempts so a genuinely stuck interface cannot hang the suite. An implicit timing assumption is thereby replaced by an explicit synchronization on observable UI state, the canonical repair for this class of flakiness~\cite{romano2021, lam2020}.

This example also exposes why detecting such defects is hard in the first place. The two outcomes are separated only by sub-second differences in animation timing, so the failure is not reproducible on demand: confirming it dynamically would mean rerunning the test many times and hoping to land on the unlucky schedule, and in UI-driven suites, where each run drives a full application through its interface, that cost is multiplied by long execution times and large input spaces~\cite{romano2021}. The very property that makes this flakiness expensive to trigger, its dependence on runtime timing, also makes it expensive to confirm by running the test.

Yet nothing about the underlying cause is hidden at rest. The entire episode, both the defect and its repair, is legible \emph{statically}: the test is saturated with UI locators (\textit{staticTexts}, \textit{accessibilityidentifiers}) and explicit synchronization primitives (\textit{waitAndTap}, \textit{waitForNoExistence}), and the fix does nothing but add more of the same vocabulary. The timing assumption that caused the flakiness, and the synchronization that removed it, are both written into the source. This is the tension our work is built on: a fault whose \emph{behaviour} only manifests at runtime nonetheless leaves a stable \emph{lexical} trace in the test code, one that is available without executing the suite.

That trace, however, is expressed through constructs specific to Swift and \textit{XCUITest}, an ecosystem on which the vocabulary-based approach to flakiness has never been evaluated (Section~\ref{sec:related-work}). 
Whether the iOS-flavoured vocabulary illustrated here is idiosyncratic to this one test or a recurring, learnable signal across real Swift projects is precisely what motivates the study that follows.

\section{Related Work}
\label{sec:related-work}

\textit{Flaky tests.} A \textit{flaky test} produces different outcomes across executions of the same version of the software, without any change to the implementation~\cite{luo2014}. Its cost is well documented: developers spend considerable effort triaging failures that signal no real defect~\cite{miranda2020}, and repeated exposure erodes trust to the point that genuine failures are ignored, harming software stability~\cite{Rahman2018}; surveys of practitioners confirm that flakiness is a recurring, everyday concern rather than an occasional nuisance~\cite{gruber2022}.

\textit{Detecting flaky tests.} Detection techniques are commonly classified as dynamic, hybrid, or static, according to whether the test must be executed~\cite{miranda2020, soratto2023}. The dynamic approach simply reruns the suite until a test disagrees with itself~\cite{parry2021}, a strategy deployed at scale at Google and Microsoft~\cite{micco2017, lam2020}; its cost, however, grows with the rerun budget, and even five reruns surface only about 88\% of \textit{flaky} tests~\cite{lam_darko_2020}. Hybrid techniques reduce this cost by pairing a single execution with additional analysis: \textit{DeFlaker} flags a failing test as \textit{flaky} when it does not exercise recently modified code, matching the detection rate of repeated re-execution at a fraction of the overhead~\cite{bell2018}, while \textit{Shaker} injects CPU and memory stress across test runs to surface concurrency-related \textit{flakiness} more quickly~\cite{denini2020}. Even so, hybrid methods still require executing the suite at least once, which remains prohibitive for large projects.

\textit{Vocabulary-based prediction.} This cost motivates \textit{static} prediction, which infers \textit{flakiness} from source code alone. \citet{miranda2020} observed that \textit{flaky} tests share a characteristic \textit{vocabulary}, identifiers and keywords that recur disproportionately, and trained classifiers such as \textit{Random Forest} and \textit{SVM} that reached an F1-score of 0.95 on a Java dataset, with the most informative tokens (\textit{job}, \textit{table}, \textit{id}, \textit{action}) pointing to root causes such as remote task execution and event queues. Subsequent work showed the signal to be robust across feature sets and classifiers, whether combining lexical with behavioral features (\textit{FlakeFlagger}~\cite{alshammari2021}) or applying nearest-neighbor search over code vectors (\textit{FLAST}~\cite{miranda2021}), and replications extended it beyond Java to JavaScript~\cite{soratto2023} and across five languages including Python, C++, and Go~\cite{ahmad2025}. \citet{barbosa2022test} further show, via re-execution across languages, that root causes are only partially shared and vary by ecosystem, evidence that language matters. Crucially, none of these studies covers Swift or any language with a UI-test-dominated suite.

\textit{Learned features.} A parallel line of work replaces hand-crafted vocabulary with pretrained language models, ranging from fine-tuning CodeBERT on test code, with reported F1-scores between 0.79 and 0.98~\cite{fatima2022flakify}, to prompting general-purpose LLMs, which in a recent evaluation performed only marginally above random guessing on test code alone~\cite{berndt2026can}. The evidence here is mixed: learned features help most for root-cause categorization and far less for the binary flaky/non-flaky prediction addressed in this work~\cite{rahman2024quantizing}, and once class imbalance is handled realistically their reported accuracy drops sharply (F1 from 0.82 to 0.57), with models leaning on superficial tokens rather than code semantics~\cite{rahman2025understanding}. Vocabulary-based features, by contrast, remain computationally cheap and directly interpretable, supporting the kind of root-cause analysis pursued in this work.

\textit{Mobile and UI testing.} Finally, studies of mobile and UI-driven suites, the dominant testing style in the Swift/iOS ecosystem, report root causes, and presumably vocabularies, that differ from those of JVM and web code: \citet{thorve2018} identify mobile-specific causes in Android not previously reported for non-mobile software, and \citet{romano2021}, analyzing 235 \textit{flaky} UI tests from 62 web and Android projects, show that the large input spaces of UI tests make repeated re-execution especially impractical, reinforcing the case for static detection. Swift/iOS, with its \textit{async}/\textit{await} concurrency model and heavy reliance on \textit{XCTest}/\textit{XCUITest}, sits squarely in this gap: it has a markedly different concurrency model and a far stronger UI-testing focus than any ecosystem on which vocabulary-based prediction has been evaluated so far. This work closes that gap, motivating the research questions investigated next.

\section{Approach}
\label{sec:approach}

Our approach builds an interpretable \emph{Swift vocabulary of flaky tests} and uses it to predict flaky tests directly from source code. It rests on a single premise: although flakiness manifests only at runtime, it leaves a stable \emph{lexical} trace in the test code.

\begin{figure*}[ht]
    \centering
    \resizebox{\linewidth}{!}{\input{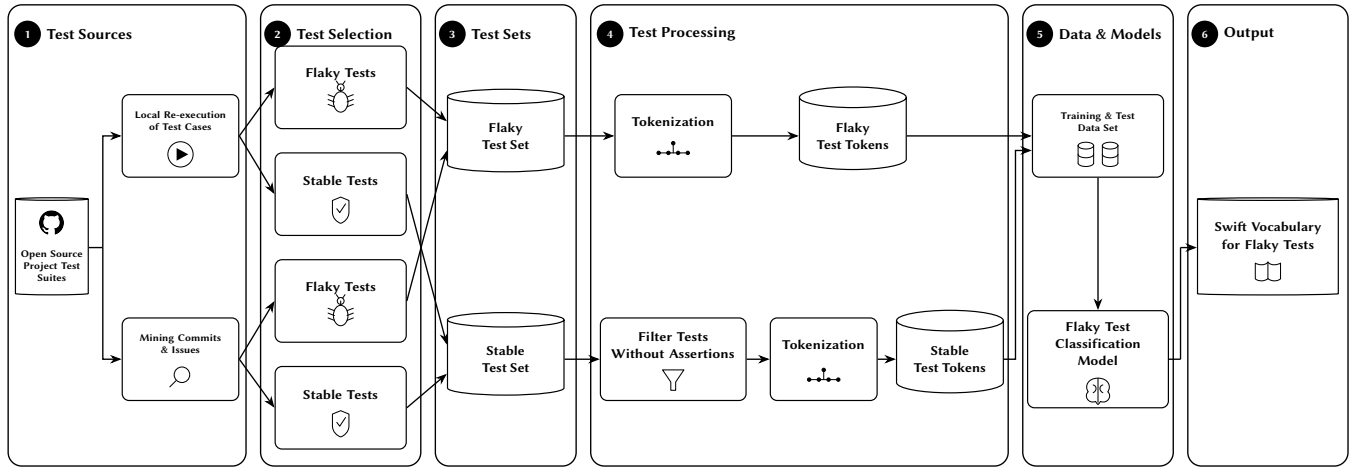}}
    \caption{End-to-end workflow for constructing the Swift vocabulary of flaky tests.}
    \label{fig:flow}
\end{figure*}

To operationalize this premise we first construct a labeled dataset from 15 open-source Swift projects (see Table~\ref{tab:subject-projects}) using two complementary sources: \emph{local re-execution} and \emph{history mining}; all remaining tests constitute the stable pool.
Figure~\ref{fig:flow} summarizes the resulting six-step workflow. We describe each step below, stressing the rationale behind each decision.

\myparagraph{Step~1: Test Sources}
Ground truth for flakiness is hard to obtain because no single oracle is at once precise and complete. We therefore draw candidate tests from open-source Swift projects (Table~\ref{tab:subject-projects}) through two complementary sources, \emph{local re-execution} and the \emph{mining of commit and pull-request histories}. The two are deliberately combined because their precision/recall trade-offs are complementary: re-execution yields high-confidence labels but only for non-determinism that reproduces in our environment, whereas mining captures a far broader range of developer-diagnosed root causes, so their union forms a more representative flaky set than either source alone.

\myparagraph{Step~2: Test Selection}
Each source is then turned into flaky and stable labels. By re-execution, we ran each test 50 times against an unchanged revision on a local macOS machine (Apple M1 Pro, 16\,GB RAM); a test that produced at least one disagreeing outcome is labeled flaky~\cite{luo2014}. By mining, we collect commits and pull requests referencing ``\textit{flaky}'' and keep only those that genuinely repair a flaky test, confirmed by manual inspection of the diff to verify that the changed method is a test and that the commit message or PR description attributes the fix to non-determinism. We then recover the test at the parent of the fixing commit, since that pre-fix version is the artifact that still exhibited the behavior (e.g., firefox-ios PR~\#27270~\cite{firefoxios_pr27270}, which fixes two flaky UI tests). Tests flagged by neither source constitute the stable pool.

\myparagraph{Step~3: Test Set Construction}
The flaky tests confirmed by the two sources are merged into a single \emph{flaky test set} and the remainder into a \emph{stable test set}. Unifying the sources is what lets one classifier learn from the full diversity of observed instability, regardless of how each case was discovered; the resulting composition and its pronounced class imbalance are detailed in Section~\ref{sec:objects}, and we assess the soundness of this hybrid labeling in Section~\ref{sec:robustness}.

\myparagraph{Step~4: Test Processing}
Each test is reduced to the lexical units that carry the signal. We first discard stable tests with no \textit{XCTest} assertion, as a method that verifies nothing is not a genuine test and would only blur the contrast between classes, and then tokenize the rest after stripping line and block comments.
We then remove 23 structural Swift keywords, spanning declarations (\textit{let}, \textit{func}, \textit{class}, \ldots) and control flow (\textit{if}, \textit{for}, \textit{guard}, \ldots); being near-ubiquitous, they would only crowd out the discriminative vocabulary. Keywords tied to concurrency or error propagation (\textit{async}, \textit{await}, \textit{throws}, \textit{try}, \textit{defer}) are deliberately retained, and the full list is in the artifact. The output is two token collections, one per class.

\myparagraph{Step~5: Model Construction}
Tests are vectorized with \textit{TF-IDF} over unigrams \emph{and} bigrams, since several instability cues are meaningful only as adjacent pairs (e.g., \textit{async throws}, \textit{load ordering}); the vectoriser keeps terms occurring in at least two documents and in at most 90\% of them, with sublinear term-frequency scaling. To keep our conclusions independent of any single learner, we train and compare five classifiers (\textit{Random Forest}, \textit{Decision Tree}, \textit{Naive Bayes}, \textit{SVM}, \textit{KNN})~\cite{miranda2020, soratto2023}. \textit{Random Forest} uses 200 trees and \textit{SVM} a linear kernel; both, with \textit{Decision Tree}, use balanced class weights, while \textit{Naive Bayes} and \textit{KNN} keep scikit-learn defaults. The five span distinct inductive biases and match those of the studies we build on, making comparison direct; with only 91 positives, fine-tuning a language model would risk overfitting and forgo the interpretability Section~\ref{sec:RQ3} depends on. Because the corpus is severely imbalanced, we undersample the majority class and evaluate under stratified 5-fold cross-validation.

\myparagraph{Step~6: Vocabulary Extraction}
Finally, we rank tokens by information gain with respect to the flaky/stable label, obtaining the \emph{Swift vocabulary of flaky tests}: the lexical units whose presence most reduces uncertainty about a test's class. This ranking is what makes the approach explanatory rather than opaque, letting the predictive signal be traced to concrete constructs and mapped onto known root causes of flakiness.

\section{Objects of Analysis}
\label{sec:objects}

The unit of analysis in this study is the individual Swift \emph{test case}, the smallest artifact for which a flaky/stable label is meaningful and the granularity at which the vocabulary signal is learned. We collected these test cases from the 15 open-source Swift projects listed in Table~\ref{tab:subject-projects}. The projects span the breadth of the Swift ecosystem relevant to our research questions: networking and server-side frameworks (\textit{Alamofire}, \textit{swift-nio}, \textit{vapor}), data and persistence libraries (\textit{GRDB.swift}, \textit{Kingfisher}, \textit{Nuke}), full iOS/macOS applications with substantial UI test suites (\textit{firefox-ios}, \textit{DuckDuckGo iOS}, \textit{Signal-iOS}, \textit{Maccy}), and core developer tooling (\textit{swift-package-manager}, \textit{swift-composable-architecture}, \textit{swift-snapshot-testing}, \textit{sentry-cocoa}). Such breadth suits our questions, exercising both the \textit{async}/\textit{await} concurrency model and the UI-testing style that distinguish Swift from the ecosystems studied in prior work.

\begin{table}[t]
\caption{Subject projects and dataset composition. Flaky tests are split by labeling source: \emph{Mined} denotes tests recovered from commit/PR histories; \emph{Re-exec.} denotes tests identified by running each test 50 times.}
\label{tab:subject-projects}
\centering
\footnotesize
\setlength{\tabcolsep}{4pt}
\begin{tabular}{lrrrrr}
\toprule
 & & \multicolumn{3}{c}{\textbf{Flaky Tests}} & \\
\cmidrule(lr){3-5}
\textbf{Project} & \textbf{\faStar} & \textbf{Mined} & \textbf{Re-exec.} & \textbf{Total} & \textbf{\# Non-Flaky} \\
\midrule
\href{https://github.com/Alamofire/Alamofire}{Alamofire}                                              & 42.4k        & 1  & 2  &  3 &  554 \\
\href{https://github.com/duckduckgo/iOS}{DuckDuckGo iOS}                                              &  1.9k        & 0  & 17 & 17 & 1003 \\
\href{https://github.com/mozilla-mobile/firefox-ios}{firefox-ios}                                     &   13k        & 8  &  1 &  9 & 4487 \\
\href{https://github.com/groue/GRDB.swift}{GRDB.swift}                                               &  8.5k        & 9  &  0 &  9 &  172 \\
\href{https://github.com/kickstarter/ios-oss}{ios-oss}                                               & 8.7k     & 1  &  0 &  1 & 1235 \\
\href{https://github.com/onevcat/Kingfisher}{Kingfisher}                                             & 24.3k        & 2  &  1 &  3 &  703 \\
\href{https://github.com/p0deje/Maccy}{Maccy}                                                        & 20.3k        & 2  &  0 &  2 &   32 \\
\href{https://github.com/kean/Nuke}{Nuke}                                                            &  8.6k        & 11 &  0 & 11 &  126 \\
\href{https://github.com/getsentry/sentry-cocoa}{sentry-cocoa}                                       &  1.1k        & 3  &  0 &  3 & 5041 \\
\href{https://github.com/signalapp/Signal-iOS}{Signal-iOS}                                           & 12.1k        & 7  &  0 &  7 &  713 \\
\href{https://github.com/pointfreeco/swift-composable-architecture}{swift-composable-architecture}   & 14.7k        & 2  &  0 &  2 &  686 \\
\href{https://github.com/swiftlang/swift-package-manager}{swift-package-manager}                     & 10.2k        & 9  &  0 &  9 & 1173 \\
\href{https://github.com/apple/swift-nio}{swift-nio}                                                 &  8.5k        & 10 &  0 & 10 & 2657 \\
\href{https://github.com/pointfreeco/swift-snapshot-testing}{swift-snapshot-testing}                 &  4.3k        & 0  &  2 &  2 & --$^\ast$ \\
\href{https://github.com/vapor/vapor}{vapor}                                                         & 26.1k        & 1  &  2 &  3 & 3767 \\
\midrule
\textbf{Total} & -- & \textbf{66} & \textbf{25} & \textbf{91} & \textbf{22,349} \\
\bottomrule
\multicolumn{6}{l}{\footnotesize $^\ast$ Suite not covered by the stable-test extraction.}\\
\end{tabular}
\end{table}

Applying the labeling procedure of Section~\ref{sec:approach} to these projects yields \textbf{91 flaky} and \textbf{22,349 non-flaky} test cases. Two properties of this dataset shape the evaluation that follows. First, the distribution is \emph{severely imbalanced}, with flaky tests accounting for roughly 0.4\% of all tests; this mirrors the reality of production suites and directly motivates both the undersampling strategy of Step~5 and our use of imbalance-aware metrics. Second, as shown in Table~\ref{tab:subject-projects}, 66 of the 91 flaky tests were obtained through history mining and 25 through local re-execution, with some projects contributing flaky tests from both sources. This hybrid composition lets us study whether the labeling strategy is sound by re-running the analysis on the re-execution subset alone.

\section{Evaluation}
\label{sec:evaluation}

We evaluate whether vocabulary-based prediction of flaky tests is effective in the Swift ecosystem, whether its results reflect a genuine lexical signal, and what that signal reveals about Swift-specific root causes of flakiness. We structure this investigation around four research questions, the first of which comprises two sub-questions addressing predictive effectiveness and labeling robustness.

\noindent\textbf{RQ1.1 --- Predictive Effectiveness.} \RQOnea

\noindent\textit{Rationale.} Vocabulary-based prediction has been shown to work for Java and JavaScript, but never for a language with the \textit{async}/\textit{await} concurrency model and UI-test focus of Swift. RQ1.1 establishes whether the lexical trace of flakiness is strong enough, in this ecosystem, for standard classifiers to separate flaky from stable tests using static features alone.

\smallskip
\noindent\textbf{RQ1.2 --- Robustness of the Hybrid Labeling Strategy.} \RQOneb

\noindent\textit{Rationale.} The conclusions of RQ1.1 are only meaningful if the flaky labels are reliable. Most labels come from commit and pull-request mining, which relies on developers explicitly documenting non-determinism and may admit noise. RQ1.2 stress-tests the labeling strategy by replicating the RQ1.1 evaluation exclusively on the 25 tests whose flakiness was confirmed by local re-execution, where the label is unambiguous.

\smallskip
\noindent\textbf{RQ2 --- Comparison with Trivial Baselines.} \RQTwo

\noindent\textit{Rationale.} On a balanced two-class problem, even naive predictors achieve non-zero scores, so high absolute metrics do not, by themselves, demonstrate that the model has learned a meaningful signal. RQ2 contextualizes the results of RQ1.1 against trivial baselines, ensuring that the observed performance reflects genuine discriminative power rather than an artifact of the evaluation setup.

\smallskip
\noindent\textbf{RQ3 --- The Vocabulary of Flaky Swift Tests.} \RQThree

\noindent\textit{Rationale.} Beyond predictive accuracy, a central goal of this work is interpretability. RQ3 asks which lexical units drive the prediction, so that the signal can be traced to concrete Swift/iOS constructs and mapped onto known root causes of flakiness, and compared with the vocabularies reported for other languages.

\smallskip
\noindent\textbf{RQ4 --- Error Analysis.} \RQFour

\noindent\textit{Rationale.} Aggregate metrics hide the conditions under which a vocabulary-based model breaks down. RQ4 examines representative misclassifications qualitatively, to understand the limits of lexical prediction and to inform when a static, vocabulary-based detector should and should not be trusted.

\subsection{Experimental Setup}
\label{sec:setup}

All research questions are evaluated under the protocol of Step~5 (Section~\ref{sec:approach}): stratified 5-fold cross-validation over class-balanced data obtained by random undersampling of the majority class. Three properties determine how the numbers should be read.

First, undersampling is applied to the corpus as a whole, so the \emph{test} folds are balanced at roughly 1:1 rather than left at the natural 0.4\% rate. Every Precision reported here is therefore measured on an artificial class ratio, not the one a detector would face in production; MCC and AUC are the metrics to compare across settings. Second, a single undersample is not a stable basis for reporting, since which 91 stable tests are drawn moves F1 by several points on its own: we repeat the draw 30 times and report the mean over the resulting 150 fold-level estimates, with a 95\% confidence interval. Third, \textit{TF-IDF} is fitted on the training fold alone, so no information from held-out tests reaches the feature space, and information gain serves only interpretation (RQ3) and the vocabulary-threshold baseline (RQ2), never feature selection: training always uses the full \textit{TF-IDF} space.

To avoid the well-known pitfalls of accuracy on imbalanced data, we report five complementary metrics, namely \emph{Precision}, \emph{Recall}, \emph{F1-Score}, the \emph{Matthews correlation coefficient} (MCC), and the area under the ROC curve (AUC), placing particular emphasis on F1, MCC, and AUC, which remain informative when the classes are unequally represented and which prior vocabulary-based studies adopt as primary metrics~\cite{miranda2020, soratto2023}. Confusion matrices are additionally reported per classifier to expose the balance between false positives and false negatives.

\subsection{Answering RQ1.1: Predictive Effectiveness}
\label{sec:RQ1}

\begin{table}[ht]
\small
\caption{Classifier results, averaged over 30 draws $\times$ 5 folds; 95\% CI at most $\pm$0.02.}
\label{tab:resultados_classificador}
\begin{tabular}{lccccc}
\toprule
\textbf{Algorithm} & \textbf{Precision} & \textbf{Recall} & \textbf{F1} & \textbf{MCC} & \textbf{AUC} \\
\midrule
Random Forest & 0.92 & 0.82 & 0.86 & 0.75 & 0.95 \\
Decision Tree & 0.82 & 0.74 & 0.77 & 0.59 & 0.79 \\
Naive Bayes & 0.81 & 0.91 & 0.85 & 0.70 & 0.94 \\
Support Vector Machine & 0.86 & 0.89 & 0.87 & 0.74 & 0.94 \\
K-Nearest Neighbours & 0.80 & 0.86 & 0.82 & 0.65 & 0.89 \\
\bottomrule
\end{tabular}
\end{table}

Under the protocol of Section~\ref{sec:setup}, all five classifiers separate flaky from non-flaky tests well (Table~\ref{tab:resultados_classificador}). \textit{Random Forest} and \textit{SVM} lead and are statistically indistinguishable from one another: \textit{Random Forest} attains the highest Precision (0.92), MCC (0.75) and AUC (0.95), while \textit{SVM} attains a marginally higher F1 (0.87 against 0.86), a difference well inside the confidence intervals of both. \textit{Naive Bayes} is competitive (F1 0.85) and buys the highest Recall of the five (0.91) at the cost of Precision, \textit{KNN} follows (F1 0.82), and \textit{Decision Tree} trails (F1 0.77). That learners with such different inductive biases agree indicates the signal is carried by the vocabulary itself rather than by any one model: Swift's concurrency idioms (\textit{async}/\textit{await}, \textit{GCD} dispatch, \textit{XCTestExpectation}) are lexically concentrated and recur across flaky tests, yielding a feature space separable even for simple classifiers.

These results indicate that the models can predict \textit{flaky tests} with confidence from attributes extracted statically from the source code. The average confusion matrices (Figure~\ref{fig:confusion_matrices}) show the trade-off each learner strikes: \textit{Random Forest} is the most conservative, averaging only 1.5 false positives per fold against 3.2 false negatives, whereas \textit{Naive Bayes} inverts that balance (4.1 against 1.6) and \textit{SVM} sits between the two. Which of these is preferable depends on the cost of a missed flaky test relative to that of a spurious warning. These findings are in agreement with \cite{miranda2020}, which demonstrated that \textit{Random Forest} and \textit{SVM} tend to present greater robustness in flaky-test detection scenarios.

\begin{tcolorbox}[answerbox]
\textbf{RQ\textsubscript{1.1}}: All five classifiers reliably distinguish flaky from non-flaky Swift tests using only static vocabulary features. \textit{Random Forest} achieved the best overall performance (Precision = 0.92, F1 = 0.86, MCC = 0.75, AUC = 0.95), with \textit{SVM} statistically indistinguishable from it.
\end{tcolorbox}

\begin{figure*}[t]
\centering\small
\begin{tabular}{@{}r|c|c|@{}}
\multicolumn{3}{c}{\textbf{Random Forest}}\\[1pt]
 & \scriptsize P:NF & \scriptsize P:F \\\hline
\scriptsize A:NF & \cellcolor{blue!12}\textbf{16.7} & \cellcolor{red!10}1.5 \\\hline
\scriptsize A:F  & \cellcolor{red!10}3.2            & \cellcolor{blue!12}\textbf{15.0}\\\hline
\end{tabular}\hfill
\begin{tabular}{@{}r|c|c|@{}}
\multicolumn{3}{c}{\textbf{Decision Tree}}\\[1pt]
 & \scriptsize P:NF & \scriptsize P:F \\\hline
\scriptsize A:NF & \cellcolor{blue!12}\textbf{15.2} & \cellcolor{red!10}3.0 \\\hline
\scriptsize A:F  & \cellcolor{red!10}4.7            & \cellcolor{blue!12}\textbf{13.5}\\\hline
\end{tabular}\hfill
\begin{tabular}{@{}r|c|c|@{}}
\multicolumn{3}{c}{\textbf{KNN}}\\[1pt]
 & \scriptsize P:NF & \scriptsize P:F \\\hline
\scriptsize A:NF & \cellcolor{blue!12}\textbf{14.1} & \cellcolor{red!10}4.1 \\\hline
\scriptsize A:F  & \cellcolor{red!10}2.6            & \cellcolor{blue!12}\textbf{15.6}\\\hline
\end{tabular}\hfill
\begin{tabular}{@{}r|c|c|@{}}
\multicolumn{3}{c}{\textbf{Naive Bayes}}\\[1pt]
 & \scriptsize P:NF & \scriptsize P:F \\\hline
\scriptsize A:NF & \cellcolor{blue!12}\textbf{14.1} & \cellcolor{red!10}4.1 \\\hline
\scriptsize A:F  & \cellcolor{red!10}1.6            & \cellcolor{blue!12}\textbf{16.6}\\\hline
\end{tabular}\hfill
\begin{tabular}{@{}r|c|c|@{}}
\multicolumn{3}{c}{\textbf{SVM}}\\[1pt]
 & \scriptsize P:NF & \scriptsize P:F \\\hline
\scriptsize A:NF & \cellcolor{blue!12}\textbf{15.4} & \cellcolor{red!10}2.8 \\\hline
\scriptsize A:F  & \cellcolor{red!10}1.9            & \cellcolor{blue!12}\textbf{16.3}\\\hline
\end{tabular}
\caption{Mean confusion matrices per fold over 30 draws $\times$ 5 folds. P\,=\,Predicted, A\,=\,Actual, NF\,=\,Non-Flaky, F\,=\,Flaky.}
\label{fig:confusion_matrices}
\end{figure*}

\subsection{Answering RQ1.2: Strategy Robustness}
\label{sec:robustness}

The results of RQ1.1 rest on a hybrid dataset in which 66 of 91 flaky tests were sourced from commit and pull-request mining and only 25 from local re-execution. The commit-mined labels depend on developers explicitly describing a fix as resolving non-determinism, which may admit noise: a commit that references ``flaky'' might fix a dependency rather than true non-determinism, or a flaky test might be retired rather than repaired. To assess whether this affects the conclusions of RQ1.1, we replicate the full evaluation using only the 25 re-execution-confirmed flaky tests, for which the label is unambiguous. The metrics obtained are presented in Table \ref{tab:resultados_reexecucao_apenas}. Every classifier stays within four percentage points of its RQ1.1 F1, and none degrades in a way that would call the labels into question: \textit{Random Forest} moves from 0.86 to 0.84, \textit{SVM} from 0.87 to 0.86, \textit{Naive Bayes} is unchanged at 0.85, \textit{KNN} improves from 0.82 to 0.81, and \textit{Decision Tree} from 0.77 to 0.81. \textit{Random Forest} becomes markedly more conservative on the smaller subset, trading Recall (0.76 against 0.82) for near-perfect Precision (0.97) and the highest AUC observed anywhere in this study (0.98).

These results support the robustness of the hybrid approach. If the commit-mined labels were substantially noisy, restricting the positive class to the 25 unambiguous re-execution cases should have \emph{improved} the metrics sharply, and training on the noisy superset should have degraded them; neither happens. The signal learned from the mined tests is therefore consistent with the signal learned from the re-executed ones. What the subset does cost is coverage: with 25 positives the confidence intervals roughly double (up to $\pm$0.04), which is why we base the main analysis on the full set.

\begin{table}
\small
\caption{Classifier results on the re-execution subset.}
\label{tab:resultados_reexecucao_apenas}
\begin{tabular}{lccccc}
\toprule
\textbf{Algorithm} & \textbf{Precision} & \textbf{Recall} & \textbf{F1} & \textbf{MCC} & \textbf{AUC} \\
\midrule
Random Forest & 0.97 & 0.76 & 0.84 & 0.76 & 0.98 \\
Decision Tree & 0.88 & 0.79 & 0.81 & 0.67 & 0.82 \\
Support Vector Machine & 0.88 & 0.87 & 0.86 & 0.75 & 0.95 \\
Naive Bayes & 0.84 & 0.87 & 0.85 & 0.71 & 0.94 \\
K-Nearest Neighbours & 0.81 & 0.83 & 0.81 & 0.63 & 0.90 \\
\bottomrule
\end{tabular}
\end{table}

\begin{tcolorbox}[answerbox]
\textbf{RQ\textsubscript{1.2}}: All five classifiers are robust to the labeling strategy, staying within four percentage points of their RQ1.1 F1 when the positive class is restricted to the 25 re-execution-confirmed tests (\textit{Random Forest}: 0.84; \textit{SVM}: 0.86; \textit{Naive Bayes}: 0.85; \textit{Decision Tree} and \textit{KNN}: 0.81). Restricting to unambiguous labels neither improves nor degrades the results appreciably, which is what one expects if the commit-mined labels are sound.
\end{tcolorbox}

\subsection{Answering RQ2: Trivial Baselines}
\label{sec:RQ2}

We compare the best-performing classifier against three trivial baselines under the same protocol: a \emph{Majority-Class Predictor} (always predicts the dominant class), a \emph{Random Predictor} (draws from the class distribution), and a \emph{Vocabulary Threshold} (predicts flaky if the test contains any of the top-10 tokens by mutual information on the training fold). The first two measure what is attainable without any vocabulary signal; the third isolates the contribution of \emph{learning} over a simple presence rule. Table~\ref{tab:baselines} reports the comparison.

\begin{table}
\small
\caption{Comparison of the best vocabulary-based classifier against trivial baselines, under the same evaluation protocol.}
\label{tab:baselines}
\begin{tabular}{lccccc}
\toprule
\textbf{Classifier} & \textbf{Precision} & \textbf{Recall} & \textbf{F1} & \textbf{MCC} & \textbf{AUC} \\
\midrule
Majority-Class Predictor  & 0.10 & 0.20 & 0.13 & 0.00 & 0.50 \\
Random Predictor          & 0.48 & 0.49 & 0.49 & $-$0.04 & 0.48 \\
Vocab.\ Threshold (top-10) & 0.54 & 0.81 & 0.64 & 0.08 & 0.52 \\
\midrule
Random Forest             & 0.92 & 0.82 & 0.86 & 0.75 & 0.95 \\
\bottomrule
\end{tabular}
\end{table}

The three baselines reveal complementary failure modes. Because undersampling makes the two classes exactly equal in size, the Majority-Class Predictor is degenerate: it has no majority to fall back on and its label depends on how the tie is broken in each fold, which is why its Precision, Recall and F1 are unstable and meaningless. Its MCC of 0.00 and AUC of 0.50 are the informative entries, and they confirm that no discriminative information flows from the label distribution alone. The Random Predictor adds label variation but likewise stays at chance (MCC = $-$0.04, AUC = 0.48).

The Vocabulary Threshold is the most diagnostic baseline: by construction, it exploits the exact tokens identified in RQ3 as most informative, the same tokens the learned model uses as features, yet it achieves an MCC of only 0.08 and an AUC of 0.52, barely above chance. The high Recall (0.81) paired with near-random Precision (0.54) reveals why. The tokens most associated with flakiness are frequent enough in the stable class that mere presence carries almost no information: a rule that flags any test containing one of them ends up flagging most of the corpus. Knowing that a test \emph{contains} a ``flaky'' token is insufficient; what matters is the weighted co-occurrence profile across the full vocabulary, which is precisely what TF-IDF combined with Random Forest learns. Raising the threshold to the top-20 tokens does not help (MCC 0.09), confirming that the limitation is the presence rule itself rather than the size of the token set.

Against this backdrop, Random Forest's MCC of 0.75 and AUC of 0.95 stand well clear of every baseline, improving MCC roughly ninefold and AUC by 0.43 points: the model does not merely detect the presence of flaky vocabulary; it learns the discriminative pattern \emph{within} that vocabulary.

\begin{tcolorbox}[answerbox]
\textbf{RQ\textsubscript{2}}: Random Forest substantially outperforms all trivial baselines, including a Vocabulary Threshold that directly exploits the top-10 mutual-information tokens yet achieves only MCC~=~0.08. The gap confirms that the model captures a genuine lexical signal that token presence alone cannot replicate.
\end{tcolorbox}

\subsection{Answering RQ3: Flakiness in Swift}
\label{sec:RQ3}

Table~\ref{tab:ganho_info_table} presents the twenty tokens with the highest information gain together with their \emph{document frequency} in each class: the percentage of tests in which the token appears at least once. Document frequencies are essential context because information gain measures discriminative power in \emph{either direction}, a token is highly informative whether it appears mostly in flaky tests (\emph{flakiness marker}) or mostly in stable tests (\emph{stability marker}, denoted~$^\dagger$). Distinguishing these two roles is critical for correct interpretation and is a finding not reported by prior vocabulary studies for other languages.

\begin{table}
\centering
\footnotesize
\setlength{\tabcolsep}{5pt}
\caption{Top-20 tokens by information gain (\textit{IG}), averaged over the 30
undersampling draws of RQ1.1. \%\,F and \%\,S: share of flaky and stable tests
containing the token at least once.}
\label{tab:ganho_info_table}
\begin{tabular}{lrrr@{\hspace{1em}}lrrr}
\toprule
\textbf{Token} & \textbf{IG} & \textbf{\%\,F} & \textbf{\%\,S} &
\textbf{Token} & \textbf{IG} & \textbf{\%\,F} & \textbf{\%\,S} \\
\midrule
\textit{xctassertequal}$^\dagger$      & 0.193 & 60 & 80 & \textit{relaxed}$^\dagger$          & 0.083 &  1 & 25 \\
\textit{async}                         & 0.142 & 38 &  1 & \textit{ordering relaxed}$^\dagger$ & 0.078 &  1 & 25 \\
\textit{throws}                        & 0.137 & 40 &  1 & \textit{try}                        & 0.071 & 44 & 17 \\
\textit{await}                         & 0.109 & 30 &  1 & \textit{expectation}                & 0.070 & 29 &  4 \\
\textit{destroy}$^\dagger$             & 0.105 &  0 & 26 & \textit{wait}                       & 0.070 & 34 &  6 \\
\textit{timeout}                       & 0.104 & 42 &  5 & \textit{description}                & 0.069 & 32 &  6 \\
\textit{xctassertequal load}$^\dagger$ & 0.102 &  0 & 26 & \textit{now}                        & 0.065 & 20 &  2 \\
\textit{unsafeatomic}$^\dagger$        & 0.098 &  0 & 26 & \textit{ordering}$^\dagger$         & 0.064 &  2 & 26 \\
\textit{defer destroy}$^\dagger$       & 0.091 &  0 & 26 & \textit{fulfill}                    & 0.062 & 27 &  5 \\
\textit{load ordering}$^\dagger$       & 0.084 &  2 & 26 & \textit{async throws}               & 0.060 & 14 &  0 \\
\midrule
\multicolumn{8}{l}{$^\dagger$ Stability marker: \%\,S\,$>$\,\%\,F.}\\
\bottomrule
\end{tabular}
\end{table}

\myparagraph{Flakiness markers}
The tokens that appear predominantly or almost exclusively in flaky tests are the strongest positive predictors of instability. \textit{Async} (38\% flaky, 1\% stable) and \textit{await} (30\% vs.\ 1\%) are the clearest examples: each is present in roughly one in three flaky tests and in almost none of the stable ones, making them the single strongest lexical signal. Their meaning is direct: tests that call \textit{async}/\textit{await} functions without sufficient synchronisation control race against the scheduler and produce non-deterministic outcomes. The bigram \textit{async throws} (14\% vs.\ 0\%) sharpens the same signal, picking out asynchronous calls that can also fail.

A second group is associated with \emph{error propagation and assertion patterns in asynchronous code}. \textit{Throws} (40\% vs.\ 1\%) appears in two out of five flaky tests and almost never in stable ones, indicating tests whose correctness depends on error-throwing async calls that can silently mis-sequence, and \textit{try} (44\% vs.\ 17\%) accompanies it at the call site. \textit{Expectation} (29\% vs.\ 4\%) and \textit{fulfill} (27\% vs.\ 5\%) represent \textit{XCTestExpectation}-based synchronisation, a mechanism specifically designed to wait for asynchronous operations: their prevalence in flaky tests indicates that expectation-based synchronisation, when misconfigured, is itself a source of non-determinism.

A third group is explicitly temporal. \textit{Timeout} (42\% vs.\ 5\%) and \textit{wait} (34\% vs.\ 6\%) mark the bounded waits through which a test states how long it is prepared to tolerate an asynchronous operation, and \textit{now} (20\% vs.\ 2\%) marks a dependence on the current clock. Together they describe tests whose outcome is contingent on execution timing, one of the most frequently reported root causes of flakiness~\cite{luo2014, eck2019}.

\myparagraph{Stability markers}
A key finding of this analysis, one not reported by prior vocabulary studies,
is the presence of \emph{stability markers}: tokens that appear more often in stable tests than in flaky ones and whose presence the model uses as evidence \emph{against} flakiness. Half of the top-20 falls into this category, and it divides into two groups of quite different character.

The first is a tightly coupled family of atomics vocabulary: \textit{unsafeatomic}, \textit{destroy}, \textit{ordering}, \textit{relaxed} and the bigrams built from them, each appearing in about a quarter of the stable tests and in essentially none of the flaky ones. These are the lexical signature of Swift's \texttt{UnsafeAtomic} API: \textit{ordering} and \textit{relaxed} name the memory-ordering qualifier passed to an atomic operation, and \textit{destroy} the paired deallocation. Tests that manipulate shared state through this API declare their synchronisation explicitly at every access, and in our corpus they are uniformly stable. The signal is strong but narrow, and Section~\ref{sec:ameacas-a-validade} examines how far it generalises.

The second group is broader. \textit{Xctassertequal} is the single most informative token in the whole ranking (IG 0.193) and it is a stability marker: it appears in 60\% of flaky tests but 80\% of stable ones. Although present in the majority of \emph{both} classes, it is more prevalent in stable tests, so the model uses it as mild evidence against flakiness. The underlying pattern is that tests saturated with straightforward equality assertions but no concurrent or asynchronous primitives are predominantly stable. It is not the assertion itself that signals instability, but the \emph{absence} of such assertions in tests that instead rely on \textit{expectation} and \textit{fulfill} for asynchronous verification.

\myparagraph{Category structure}
The top-20 tokens in this dataset organise into four groups: (i)~\emph{concurrency and asynchrony} (\textit{async}, \textit{await}, \textit{async throws}, \textit{expectation}, \textit{fulfill}), the dominant flakiness signal; (ii)~\emph{async error propagation} (\textit{throws}, \textit{try}); (iii)~\emph{timing dependence} (\textit{timeout}, \textit{wait}, \textit{now}), patterns specific to testing asynchronous code; and (iv)~\emph{stability markers}, comprising the assertion vocabulary of straightforwardly synchronous tests (\textit{xctassertequal}) and the explicit memory-ordering vocabulary of atomic operations (\textit{unsafeatomic}, \textit{ordering}, \textit{relaxed}). UI interaction tokens (e.g., \textit{accessibilityidentifiers}), while present in the dataset, do not appear among the top-20 discriminative features; this is consistent with the project composition, where 60\% of the flaky tests (55 of 91) originate from projects related to networking/server-side, persistence, and developer tooling, rather than UI-heavy applications, so concurrency vocabulary dominates the signal.

\begin{tcolorbox}[answerbox]
\textbf{RQ\textsubscript{3}}: The top-20 tokens divide into two kinds. \emph{Flakiness markers}, which appear predominantly in flaky tests, comprise concurrency primitives (\textit{async}, \textit{await}, \textit{async throws}), expectation-based synchronisation (\textit{expectation}, \textit{fulfill}), error propagation (\textit{throws}, \textit{try}), and explicit timing dependence (\textit{timeout}, \textit{wait}, \textit{now}). \emph{Stability markers} are evidence \emph{against} flakiness: chiefly \textit{xctassertequal}, the assertion vocabulary of plainly synchronous tests, alongside the explicit memory-ordering vocabulary of atomic operations. This bidirectional discriminative structure reflects both the constructs that introduce non-determinism and those that accompany its absence.
\end{tcolorbox}

\subsection{Answering RQ4: Error Analysis}
\label{sec:RQ4}

The aggregate metrics of RQ1.1 and the vocabulary of RQ3 establish \emph{that} lexical prediction works and \emph{which} tokens drive it, but they do not reveal \emph{when} it fails. To understand the limits of vocabulary-based prediction, we conduct a qualitative analysis of the misclassifications produced by the best model (\textit{Random Forest}). A complete cross-validation run produces, on average, 16.0~false negatives and 7.4~false positives. Because any single undersample yields a slightly different error set, we do not read individual errors off one run: instead we record, for every test, the fraction of the 150 models (30 draws $\times$ 5 folds) that flag it, and treat a test as a systematic error only when the majority of models that never saw it during training get it wrong. Thirteen of the 91 flaky tests are missed under this criterion. We group the resulting errors into three recurring patterns that together delimit where a purely lexical signal is and is not sufficient.

\myparagraph{False negatives: flaky tests without common flaky vocabulary}
Some genuinely flaky tests are missed because their source does not contain the tokens most associated with instability (Table~\ref{tab:ganho_info_table}). These are cases where the non-determinism originates outside the lexically visible test body, for example in a shared helper, a fixture, or an implicit ordering dependency, leaving no characteristic vocabulary for the model to latch onto. A concrete example is \texttt{testBuildParametersWithInvalidDevice} from Signal-iOS, which only 3\% of models flag. The test exercises cryptographic key-pair generation and mock-based message dispatch; its visible vocabulary consists of domain-specific identifiers (\textit{pniKeyPair}, \textit{localSignedPreKey}, \textit{localRegistrationId}) with no occurrence of the concurrency or timing tokens listed in Table~\ref{tab:ganho_info_table}. The non-determinism resides in shared mock state coordinated across the test class, outside the test body, a root cause structurally invisible to a lexical classifier. A second example, also never flagged, is shown in Figure~\ref{fig:rq4-examples}(a): \texttt{testFlush\_BlocksCallingThread\_TimesOut} from SentryCocoa~\cite{sentry_cocoa} asserts that a blocking call completes within a $\pm0.1$\,s window, yet its timing tokens (\textit{getAbsoluteTime}, \textit{toTimeInterval}) do not appear among the top-20 mutual-information features and are therefore unrecognized as a flakiness signal. These misses are systematic rather than incidental: six of the thirteen are flagged by no model at all.

\myparagraph{False positives: stable tests with flaky-associated vocabulary}
Conversely, some stable tests are flagged as flaky because they contain tokens that are strongly predictive in aggregate yet, in context, are benign. Figure~\ref{fig:rq4-examples}(b) shows the clearest example: \texttt{test\-DownsamplingHandleScale2x} from Kingfisher, flagged by 80\% of the models that never saw it. Its lexical profile is that of a textbook flaky test: it opens an \textit{XCTestExpectation}, issues an image retrieval whose result arrives in a closure, and blocks on \textit{waitForExpectations} with an explicit one-second timeout, a structure indistinguishable from tests that genuinely race against asynchronous image decoding. What makes it deterministic is the single call to \textit{stub}, which replaces the network with canned data so the callback fires synchronously inside the test process. The construct that guarantees stability is therefore present in the source, but it is one token among many and carries none of the discriminative weight that \textit{expectation} and \textit{waitForExpectations} do.

Notably, two constructs we expected to mislead the model do not. Tests built on a virtual clock (\texttt{testRunScheduler}, which uses \textit{DispatchQueue.test}) or on a synchronous in-process event loop (\texttt{testAndAllCompleteWithPreFailedFutures}, which uses \textit{EmbeddedEventLoop}) are flagged by only 7\% and 12\% of models respectively, and are thus classified correctly. Their deterministic testing infrastructure evidently brings its own vocabulary, which the model learns to read as evidence \emph{against} flakiness, in the same way as the stability markers of Section~\ref{sec:RQ3}.

\begin{figure*}[th]
\centering
\begin{minipage}[t]{0.47\textwidth}
\small\textbf{(a) False negative} --- \textit{testFlush\_BlocksCallingThread\_TimesOut}~\cite{sentry_cocoa}\\[3pt]
\begin{lstlisting}[style=swiftmini, frame=single, breaklines=true]
func testFlush_BlocksCallingThread_TimesOut() {
    givenCachedEvents(amount: 30)
    fixture.requestManager.responseDelay = fixture.flushTimeout + 0.2
    let before = SentryDate.getAbsoluteTime()
    let result = sut.flush(fixture.flushTimeout)
    let elapsed = getDurationNs(before, SentryDate.getAbsoluteTime()).toTimeInterval()
    XCTAssertGreaterThan(elapsed, fixture.flushTimeout)
    XCTAssertLessThan(elapsed, fixture.flushTimeout + 0.1) // tight 0.1s window
    XCTAssertEqual(.timedOut, result)
}
\end{lstlisting}
\end{minipage}
\hfill
\begin{minipage}[t]{0.47\textwidth}
\small\textbf{(b) False positive} --- \textit{testDownsamplingHandleScale2x} (Kingfisher)\\[3pt]
\begin{lstlisting}[style=swiftmini, frame=single, breaklines=true]
func testDownsamplingHandleScale2x() {
    let exp = expectation(description: #function)
    let url = testURLs[0]
    stub(url, data: testImageData)  // network stubbed
    _ = manager.retrieveImage(with: .network(url),
        options: [.processor(...), .scaleFactor(2)]) { result in
        let image = result.value?.image
        XCTAssertEqual(image?.size, .init(width: 4, height: 4))
        exp.fulfill()   // fires synchronously
    }
    waitForExpectations(timeout: 1, handler: nil)
}
\end{lstlisting}
\end{minipage}

\caption{Systematic misclassifications: (a)~a flaky test flagged by no model; (b)~a stable test flagged by 80\% of models.}
\label{fig:rq4-examples}
\end{figure*}

\myparagraph{Borderline cases}
Between the systematic errors and the confident predictions lies a band of tests the model cannot settle. \textit{SwiftNIO\_testMetricsDelegateTickInfo} and \textit{nuke\_testIsLoadingUpdated} are flagged by 43\% and 40\% of models respectively: genuinely flaky tests whose vocabulary straddles both classes, so the verdict flips with the training sample. This band is the practical reason for reporting the flag rate rather than a single label, and it is where a static detector is least useful on its own.

\myparagraph{Limits of vocabulary-based prediction}
These errors share one root: vocabulary captures the \emph{constructs} correlated with flakiness but not the \emph{correctness of their use}. The detector is thus most reliable when the test body itself instantiates the primitives that cause non-determinism (raw async/await, real network calls, shared mutable state), and least reliable when asynchrony is delegated to infrastructure whose determinism its own tokens do not reveal. The natural complements are data-flow analysis, to check whether async primitives are guarded, and cross-test dependency analysis, to surface shared fixtures.

\begin{tcolorbox}[answerbox]
\textbf{RQ\textsubscript{4}}: Vocabulary-based prediction fails when flakiness is hidden in shared infrastructure outside the test body (false negatives) or when async and timing constructs are used in a deterministic context (virtual schedulers, embedded event loops) that is lexically indistinguishable from a real one (false positives). In both cases, the model detects the relevant constructs but cannot assess the correctness of their use, a distinction that requires semantic rather than lexical analysis.
\end{tcolorbox}

\section{Discussion}
\label{sec:discussion}

The answers to RQ1.1, RQ1.2, and RQ3 are complementary: RQ1.1 shows that the vocabulary of a Swift test, captured through simple TF-IDF features, is sufficient for classifiers such as \textit{Random Forest} and \textit{SVM} to predict flakiness with high precision and AUC; RQ1.2 confirms that this holds under stricter labeling; and RQ3 shows that this predictive power is not an artifact of opaque features, but stems from a small, interpretable set of tokens that act in two directions. The \emph{flakiness markers} identified in Section~\ref{sec:RQ3}, namely concurrency primitives (\textit{async}, \textit{await}), expectation-based synchronisation (\textit{expectation}, \textit{fulfill}), error-propagation constructs (\textit{throws}, \textit{try}), and explicit timing dependence (\textit{timeout}, \textit{wait}, \textit{now}), map onto root causes of flakiness reported in the broader literature, including concurrency, time dependency, and resource management \cite{luo2014, eck2019}. The \emph{stability markers} represent a complementary signal: the vocabulary of tests that assert plainly and synchronously, which the model uses as evidence against flakiness.

These findings relate to, but are not identical to, the vocabulary reported for other languages. \citet{miranda2020} found that the tokens most strongly associated with flakiness in Java were related to remote task execution and event queues (e.g., \textit{job}, \textit{table}, \textit{id}, \textit{action}), pointing to a vocabulary centered on distributed/asynchronous infrastructure. In our Swift dataset, the concurrency-related signal is expressed through language-level primitives (\textit{async}, \textit{await}, \textit{async throws}) and asynchronous assertion patterns (\textit{expectation}, \textit{fulfill}, \textit{throws}). A distinctive feature not reported for Java is the presence of stability markers: tokens whose \emph{absence} in concurrency-heavy tests raises the flaky prediction. UI interaction tokens (\textit{accessibilityidentifiers}), while present in the corpus, do not appear among the top-20 discriminative features in this dataset; this is consistent with the project composition, where 60\% of the flaky tests (55 of 91) originate from non-UI projects (networking/server-side, persistence, and developer tooling), where concurrency vocabulary dominates. This suggests that while the underlying phenomenon (non-determinism caused by concurrency, timing, and external state) is consistent across ecosystems, its lexical manifestation is shaped by the language's idioms and the dominant testing style of the projects studied.

Two further experiments bound this. On the natural distribution (training balanced, test folds left at the real 0.4\% rate) Recall and ranking hold up (0.82, AUC 0.947) but Precision collapses to 0.041: catching 15 flaky tests costs 374 false alarms. Leave-one-project-out gives a macro-averaged F1 of 0.71 and AUC of 0.90, against 0.86 and 0.95 when a project may appear on both sides of the split, ranging from 0.91 on \textit{GRDB.swift} to 0.20 on \textit{firefox-ios}, the most UI-heavy subject; part of the headline figure is thus within-project vocabulary. A binary CI gate would therefore be unusable, but the model's \emph{ordering} is not: average precision is 0.31 against a 0.004 baseline, and half of the ten highest-scored tests are genuinely flaky. The realistic use is a prioritiser that spends a limited rerun budget where it pays off, which keeps the approach attractive for the expensive UI suites \cite{romano2021}, with weaker guidance expected on projects unlike the training corpus.

\section{Threats to Validity}
\label{sec:ameacas-a-validade}

\myparagraph{Construct validity} Flaky labels are derived from two sources with complementary noise profiles: commit and pull-request mining depends on developers explicitly documenting non-determinism and may admit both false positives and false negatives; local re-execution used a single uncontrolled macOS machine, which may have masked latent flakiness or introduced spurious failures. To assess the impact of label noise, we replicated the full evaluation using only the 25 re-execution-confirmed flaky tests, for which the label is unambiguous. Four of five classifiers remained within two percentage points of their full-dataset F1 on this stricter subset, supporting the soundness of the hybrid labeling strategy.

\myparagraph{Internal validity} The severe class imbalance required random undersampling, which also balances the test folds and so makes every Precision reported in Section~\ref{sec:evaluation} an optimistic reading of production behaviour; Section~\ref{sec:discussion} quantifies the gap (Precision 0.041 at the natural 0.4\% base rate) and the cross-project one. Repeating the undersample 30 times removes the dependence on any single draw, and stratified 5-fold cross-validation mitigates overfitting, but the small flaky-test pool inherently constrains generalization.

\myparagraph{External validity} All 15 projects are public open-source Swift repositories on GitHub. The limited availability of large Swift projects with mature automated test suites, an ecosystem constraint rather than a sampling choice, restricts project diversity. Findings may not transfer to proprietary codebases or to Swift projects in domains or testing styles substantially different from those studied here.

\myparagraph{Conclusion validity} To avoid misleading conclusions in the context of imbalanced data, we reported multiple evaluation metrics: precision, \textit{recall}, \textit{F1-Score}, Matthews correlation coefficient (\textit{MCC}), and area under the ROC curve (\textit{AUC}).

\section{Conclusion}
\label{sec:conclusion}
We presented the first empirical study of vocabulary-based static prediction of flaky tests in \textit{Swift}, a language with a native \textit{async}/\textit{await} concurrency model and a strong reliance on UI testing. On 91 flaky and 22,349 non-flaky tests from 15 open-source projects, source-code tokens alone proved effective predictors of instability: \textit{Random Forest} reached an F1-Score of 0.86 and AUC of 0.95, far above every trivial baseline (MCC\,=\,0.75 vs.\ 0.08), and the result holds when the positive class is restricted to re-execution-confirmed labels. The discriminative tokens are grounded in Swift's concurrency semantics, with \textit{async}, \textit{await}, \textit{throws}, \textit{expectation} and \textit{timeout} marking flakiness and \textit{xctassertequal}, the vocabulary of plainly synchronous tests, marking its absence. The errors delimit the approach: flakiness hidden in shared infrastructure is missed and deterministic uses of async constructs are flagged, both distinctions being semantic rather than lexical. Since precision falls sharply at the natural base rate, the practical value lies in ranking rather than gating, and closing the remaining gap will require semantic and structural signals such as data-flow-aware features and interprocedural token propagation.

\section*{Artifact Availability}
\vspace{-0.3em}
The replication package is available at \ourrepo. It contains the labeled corpus with per-test provenance, the scripts that regenerate every table reported here, and a Dockerfile that runs them with nothing installed locally. Our scripts are released under the MIT License and our curated data under CC BY 4.0; the Swift test excerpts remain under the licenses of the projects they were taken from, as recorded in the package's \texttt{LICENSE} and \texttt{NOTICE} files.

\bibliographystyle{ACM-Reference-Format}
\bibliography{biblio}


\begin{thebibliography}{28}


\ifx \showCODEN    \undefined \def \showCODEN     #1{\unskip}     \fi
\ifx \showISBNx    \undefined \def \showISBNx     #1{\unskip}     \fi
\ifx \showISBNxiii \undefined \def \showISBNxiii  #1{\unskip}     \fi
\ifx \showISSN     \undefined \def \showISSN      #1{\unskip}     \fi
\ifx \showLCCN     \undefined \def \showLCCN      #1{\unskip}     \fi
\ifx \shownote     \undefined \def \shownote      #1{#1}          \fi
\ifx \showarticletitle \undefined \def \showarticletitle #1{#1}   \fi
\ifx \showURL      \undefined \def \showURL       {\relax}        \fi
\providecommand\bibfield[2]{#2}
\providecommand\bibinfo[2]{#2}
\providecommand\natexlab[1]{#1}
\providecommand\showeprint[2][]{arXiv:#2}

\bibitem[Ahmad et~al\mbox{.}(2025)]%
        {ahmad2025}
\bibfield{author}{\bibinfo{person}{Azeem Ahmad}, \bibinfo{person}{Xin Sun},
  \bibinfo{person}{Muhammad~Rashid Naeem}, \bibinfo{person}{Yasir Javed},
  \bibinfo{person}{Mohammad Akour}, {and} \bibinfo{person}{Kristian Sandahl}.}
  \bibinfo{year}{2025}\natexlab{}.
\newblock \showarticletitle{Understanding Flaky Tests Through Linguistic
  Diversity: A Cross-Language and Comparative Machine Learning Study}.
\newblock \bibinfo{journal}{\emph{IEEE Access}}  \bibinfo{volume}{13}
  (\bibinfo{year}{2025}).
\newblock


\bibitem[Alshammari et~al\mbox{.}(2021)]%
        {alshammari2021}
\bibfield{author}{\bibinfo{person}{Abdulrahman Alshammari},
  \bibinfo{person}{Christopher Morris}, \bibinfo{person}{Michael Hilton}, {and}
  \bibinfo{person}{Jonathan Bell}.} \bibinfo{year}{2021}\natexlab{}.
\newblock \showarticletitle{FlakeFlagger: Predicting Flakiness Without
  Rerunning Tests}. In \bibinfo{booktitle}{\emph{43rd International Conference
  on Software Engineering}}.
\newblock


\bibitem[Barbosa et~al\mbox{.}(2022)]%
        {barbosa2022test}
\bibfield{author}{\bibinfo{person}{Keila Barbosa}, \bibinfo{person}{Ronivaldo
  Ferreira}, \bibinfo{person}{Gustavo Pinto}, \bibinfo{person}{Marcelo
  d'Amorim}, {and} \bibinfo{person}{Breno Miranda}.}
  \bibinfo{year}{2022}\natexlab{}.
\newblock \showarticletitle{Test flakiness across programming languages}.
\newblock \bibinfo{journal}{\emph{IEEE Transactions on Software Engineering}}
  (\bibinfo{year}{2022}).
\newblock


\bibitem[Bell et~al\mbox{.}(2018)]%
        {bell2018}
\bibfield{author}{\bibinfo{person}{Jonathan Bell}, \bibinfo{person}{Owolabi
  Legunsen}, \bibinfo{person}{Michael Hilton}, \bibinfo{person}{Lamyaa
  Eloussi}, \bibinfo{person}{Tifany Yung}, {and} \bibinfo{person}{Darko
  Marinov}.} \bibinfo{year}{2018}\natexlab{}.
\newblock \showarticletitle{DeFlaker: automatically detecting flaky tests}. In
  \bibinfo{booktitle}{\emph{40th International Conference on Software
  Engineering}}.
\newblock


\bibitem[Berndt et~al\mbox{.}(2026)]%
        {berndt2026can}
\bibfield{author}{\bibinfo{person}{Alexander Berndt}, \bibinfo{person}{Vekil
  Bekmyradov}, \bibinfo{person}{Rainer Gemulla}, \bibinfo{person}{Marcus
  Kessel}, \bibinfo{person}{Thomas Bach}, {and} \bibinfo{person}{Sebastian
  Baltes}.} \bibinfo{year}{2026}\natexlab{}.
\newblock \showarticletitle{Can We Classify Flaky Tests Using Only Test Code?
  An LLM-Based Empirical Study}.
\newblock \bibinfo{journal}{\emph{arXiv preprint arXiv:2602.05465}}
  (\bibinfo{year}{2026}).
\newblock


\bibitem[Contributors(2026)]%
        {sentry_cocoa}
\bibfield{author}{\bibinfo{person}{Sentry~Cocoa Contributors}.}
  \bibinfo{year}{2026}\natexlab{}.
\newblock
  \bibinfo{howpublished}{\url{https://github.com/getsentry/sentry-cocoa}}.
\newblock


\bibitem[Eck et~al\mbox{.}(2019)]%
        {eck2019}
\bibfield{author}{\bibinfo{person}{Moritz Eck}, \bibinfo{person}{Fabio
  Palomba}, \bibinfo{person}{Marco Castelluccio}, {and}
  \bibinfo{person}{Alberto Bacchelli}.} \bibinfo{year}{2019}\natexlab{}.
\newblock \showarticletitle{Understanding flaky tests: the developer’s
  perspective}. In \bibinfo{booktitle}{\emph{27th ACM Joint Meeting on European
  Software Engineering Conference and Symposium on the Foundations of Software
  Engineering}}.
\newblock


\bibitem[Fatima et~al\mbox{.}(2022)]%
        {fatima2022flakify}
\bibfield{author}{\bibinfo{person}{Sakina Fatima}, \bibinfo{person}{Taher~A
  Ghaleb}, {and} \bibinfo{person}{Lionel Briand}.}
  \bibinfo{year}{2022}\natexlab{}.
\newblock \showarticletitle{Flakify: A black-box, language model-based
  predictor for flaky tests}.
\newblock \bibinfo{journal}{\emph{IEEE TSE}} (\bibinfo{year}{2022}).
\newblock


\bibitem[Gruber and Fraser(2022)]%
        {gruber2022}
\bibfield{author}{\bibinfo{person}{Martin Gruber} {and} \bibinfo{person}{Gordon
  Fraser}.} \bibinfo{year}{2022}\natexlab{}.
\newblock \showarticletitle{A Survey on How Test Flakiness Affects Developers
  and What Support They Need To Address It}. In \bibinfo{booktitle}{\emph{2022
  IEEE Conference on Software Testing, Verification and Validation (ICST)}}.
  IEEE, \bibinfo{pages}{82--92}.
\newblock


\bibitem[Herzig and Nagappan(2015)]%
        {herzig2015}
\bibfield{author}{\bibinfo{person}{Kim Herzig} {and}
  \bibinfo{person}{Nachiappan Nagappan}.} \bibinfo{year}{2015}\natexlab{}.
\newblock \showarticletitle{Empirically detecting false test alarms using
  association rules}. In \bibinfo{booktitle}{\emph{37th International
  Conference on Software Engineering - Volume 2}}.
\newblock


\bibitem[iOS Contributors(2025)]%
        {firefoxios_pr27270}
\bibfield{author}{\bibinfo{person}{Mozilla iOS Contributors}.}
  \bibinfo{year}{2025}\natexlab{}.
\newblock \bibinfo{title}{Pull Request \#27270}.
\newblock
  \bibinfo{howpublished}{\url{https://github.com/mozilla-mobile/firefox-ios/pull/27270/}}.
\newblock


\bibitem[Kapfhammer(2004)]%
        {Kapfhammer2004}
\bibfield{author}{\bibinfo{person}{Gregory~M. Kapfhammer}.}
  \bibinfo{year}{2004}\natexlab{}.
\newblock \showarticletitle{Software testing}.
\newblock In \bibinfo{booktitle}{\emph{The Computer Science Handbook}}.
  \bibinfo{publisher}{CRC Press}.
\newblock


\bibitem[Lam et~al\mbox{.}(2020a)]%
        {lam2020}
\bibfield{author}{\bibinfo{person}{Wing Lam}, \bibinfo{person}{K\i{}van\c{c}
  Mu\c{s}lu}, \bibinfo{person}{Hitesh Sajnani}, {and} \bibinfo{person}{Suresh
  Thummalapenta}.} \bibinfo{year}{2020}\natexlab{a}.
\newblock \showarticletitle{A study on the lifecycle of flaky tests}. In
  \bibinfo{booktitle}{\emph{42nd International Conference on Software
  Engineering}}.
\newblock


\bibitem[Lam et~al\mbox{.}(2020b)]%
        {winter2020}
\bibfield{author}{\bibinfo{person}{Wing Lam}, \bibinfo{person}{Stefan Winter},
  \bibinfo{person}{Angello Astorga}, \bibinfo{person}{Victoria Stodden}, {and}
  \bibinfo{person}{Darko Marinov}.} \bibinfo{year}{2020}\natexlab{b}.
\newblock \showarticletitle{Understanding Reproducibility and Characteristics
  of Flaky Tests Through Test Reruns in Java Projects}. In
  \bibinfo{booktitle}{\emph{IEEE 31st International Symposium on Software
  Reliability Engineering (ISSRE)}}.
\newblock


\bibitem[Lam et~al\mbox{.}(2020c)]%
        {lam_darko_2020}
\bibfield{author}{\bibinfo{person}{Wing Lam}, \bibinfo{person}{Stefan Winter},
  \bibinfo{person}{Angello Astorga}, \bibinfo{person}{Victoria Stodden}, {and}
  \bibinfo{person}{Darko Marinov}.} \bibinfo{year}{2020}\natexlab{c}.
\newblock \showarticletitle{Understanding reproducibility and characteristics
  of flaky tests through test reruns in java projects}
  \emph{(\bibinfo{series}{Proceedings - International Symposium on Software
  Reliability Engineering, ISSRE})}. \bibinfo{publisher}{IEEE Computer
  Society}.
\newblock


\bibitem[Luo et~al\mbox{.}(2014)]%
        {luo2014}
\bibfield{author}{\bibinfo{person}{Qingzhou Luo}, \bibinfo{person}{Farah
  Hariri}, \bibinfo{person}{Lamyaa Eloussi}, {and} \bibinfo{person}{Darko
  Marinov}.} \bibinfo{year}{2014}\natexlab{}.
\newblock \showarticletitle{An empirical analysis of flaky tests}. In
  \bibinfo{booktitle}{\emph{22nd ACM SIGSOFT International Symposium on
  Foundations of Software Engineering}}.
\newblock


\bibitem[Micco(2017)]%
        {micco2017}
\bibfield{author}{\bibinfo{person}{John Micco}.}
  \bibinfo{year}{2017}\natexlab{}.
\newblock \bibinfo{title}{The State of Continuous Integration Testing @Google}.
\newblock


\bibitem[Mozilla(2026)]%
        {firefoxProj}
\bibfield{author}{\bibinfo{person}{Mozilla}.} \bibinfo{year}{2026}\natexlab{}.
\newblock \bibinfo{title}{Firefox for iOS: Open-Source Web Browser for iOS}.
\newblock
\urldef\tempurl%
\url{https://github.com/mozilla-mobile/firefox-ios}
\showURL{%
\tempurl}


\bibitem[Parry et~al\mbox{.}(2021)]%
        {parry2021}
\bibfield{author}{\bibinfo{person}{Owain Parry}, \bibinfo{person}{Gregory~M.
  Kapfhammer}, \bibinfo{person}{Michael Hilton}, {and} \bibinfo{person}{Phil
  McMinn}.} \bibinfo{year}{2021}\natexlab{}.
\newblock \showarticletitle{A Survey of Flaky Tests}.
\newblock \bibinfo{journal}{\emph{ACM Trans. Softw. Eng. Methodol.}}
  \bibinfo{volume}{31}, \bibinfo{number}{1}, Article \bibinfo{articleno}{17}
  (\bibinfo{year}{2021}).
\newblock
\showISSN{1049-331X}


\bibitem[Pinto et~al\mbox{.}(2020)]%
        {miranda2020}
\bibfield{author}{\bibinfo{person}{Gustavo Pinto}, \bibinfo{person}{Breno
  Miranda}, \bibinfo{person}{Supun Dissanayake}, \bibinfo{person}{Marcelo
  d'Amorim}, \bibinfo{person}{Christoph Treude}, {and} \bibinfo{person}{Antonia
  Bertolino}.} \bibinfo{year}{2020}\natexlab{}.
\newblock \showarticletitle{What is the Vocabulary of Flaky Tests?}. In
  \bibinfo{booktitle}{\emph{17th International Conference on Mining Software
  Repositories}}.
\newblock


\bibitem[Rahman and Rigby(2018)]%
        {Rahman2018}
\bibfield{author}{\bibinfo{person}{Md~Tajmilur Rahman} {and}
  \bibinfo{person}{Peter~C. Rigby}.} \bibinfo{year}{2018}\natexlab{}.
\newblock \showarticletitle{The impact of failing, flaky, and high failure
  tests on the number of crash reports associated with Firefox builds}. In
  \bibinfo{booktitle}{\emph{26th ACM Joint Meeting on European Software
  Engineering Conference and Symposium on the Foundations of Software
  Engineering}}.
\newblock


\bibitem[Rahman et~al\mbox{.}(2024)]%
        {rahman2024quantizing}
\bibfield{author}{\bibinfo{person}{Shanto Rahman}, \bibinfo{person}{Abdelrahman
  Baz}, \bibinfo{person}{Sasa Misailovic}, {and} \bibinfo{person}{August Shi}.}
  \bibinfo{year}{2024}\natexlab{}.
\newblock \showarticletitle{Quantizing large-language models for predicting
  flaky tests}. In \bibinfo{booktitle}{\emph{2024 IEEE Conference on Software
  Testing, Verification and Validation (ICST)}}. IEEE,
  \bibinfo{pages}{93--104}.
\newblock


\bibitem[Rahman et~al\mbox{.}(2025)]%
        {rahman2025understanding}
\bibfield{author}{\bibinfo{person}{Shanto Rahman}, \bibinfo{person}{Saikat
  Dutta}, {and} \bibinfo{person}{August Shi}.} \bibinfo{year}{2025}\natexlab{}.
\newblock \showarticletitle{Understanding and improving flaky test
  classification}.
\newblock \bibinfo{journal}{\emph{Proceedings of the ACM on Programming
  Languages}} \bibinfo{volume}{9}, \bibinfo{number}{OOPSLA2}
  (\bibinfo{year}{2025}), \bibinfo{pages}{1345--1371}.
\newblock


\bibitem[Rampim~Soratto and Graciotto~Silva(2023)]%
        {soratto2023}
\bibfield{author}{\bibinfo{person}{Rafael Rampim~Soratto} {and}
  \bibinfo{person}{Marco~Aur\'{e}lio Graciotto~Silva}.}
  \bibinfo{year}{2023}\natexlab{}.
\newblock \showarticletitle{Vocabulary of Flaky Tests in Javascript}. In
  \bibinfo{booktitle}{\emph{XXII Brazilian Symposium on Software Quality}}.
\newblock


\bibitem[Romano et~al\mbox{.}(2021)]%
        {romano2021}
\bibfield{author}{\bibinfo{person}{Alan Romano}, \bibinfo{person}{Zihe Song},
  \bibinfo{person}{Sampath Grandhi}, \bibinfo{person}{Wei Yang}, {and}
  \bibinfo{person}{Weihang Wang}.} \bibinfo{year}{2021}\natexlab{}.
\newblock \showarticletitle{An empirical analysis of UI-based flaky tests}. In
  \bibinfo{booktitle}{\emph{2021 IEEE/ACM 43rd International Conference on
  Software Engineering (ICSE)}}. IEEE, \bibinfo{pages}{1585--1597}.
\newblock


\bibitem[Silva et~al\mbox{.}(2020)]%
        {denini2020}
\bibfield{author}{\bibinfo{person}{Denini Silva}, \bibinfo{person}{Leopoldo
  Teixeira}, {and} \bibinfo{person}{Marcelo d’Amorim}.}
  \bibinfo{year}{2020}\natexlab{}.
\newblock \showarticletitle{Shake It! Detecting Flaky Tests Caused by
  Concurrency with Shaker}. In \bibinfo{booktitle}{\emph{2020 IEEE
  International Conference on Software Maintenance and Evolution (ICSME)}}.
  \bibinfo{pages}{301--311}.
\newblock


\bibitem[Thorve et~al\mbox{.}(2018)]%
        {thorve2018}
\bibfield{author}{\bibinfo{person}{Swapna Thorve}, \bibinfo{person}{Chandani
  Sreshtha}, {and} \bibinfo{person}{Na Meng}.} \bibinfo{year}{2018}\natexlab{}.
\newblock \showarticletitle{An empirical study of flaky tests in android apps}.
  In \bibinfo{booktitle}{\emph{2018 IEEE International Conference on Software
  Maintenance and Evolution (ICSME)}}. IEEE, \bibinfo{pages}{534--538}.
\newblock


\bibitem[Verdecchia et~al\mbox{.}(2021)]%
        {miranda2021}
\bibfield{author}{\bibinfo{person}{Roberto Verdecchia}, \bibinfo{person}{Emilio
  Cruciani}, \bibinfo{person}{Breno Miranda}, {and} \bibinfo{person}{Antonia
  Bertolino}.} \bibinfo{year}{2021}\natexlab{}.
\newblock \showarticletitle{Know You Neighbor: Fast Static Prediction of Test
  Flakiness}.
\newblock \bibinfo{journal}{\emph{IEEE Access}} (\bibinfo{year}{2021}).
\newblock


\end{thebibliography}

\end{document}